\documentclass[12pt]{iopart}

\expandafter\let\csname equation*\endcsname\relax
\expandafter\let\csname endequation*\endcsname\relax 

\usepackage{amsmath}
\usepackage{amssymb}
\usepackage{mathrsfs}
\usepackage{epsfig}
\usepackage{color}
\usepackage[bookmarks]{hyperref}
\usepackage{filecontents}
\usepackage{graphicx,multirow}
\usepackage{mathrsfs}
\usepackage{dsfont}
\graphicspath{{.}{./}}
\usepackage{enumerate}
\usepackage{mathtools}
\usepackage{bm}
\usepackage{float}
\usepackage{soul}

\newcommand* {\bra}[1]{\ensuremath{\langle {#1} |}}
\newcommand* {\ket}[1]{\ensuremath{| {#1} \rangle}}

\begin{document}

\title[]{Entanglement in bipartite quantum systems: Euclidean volume ratios and detectability by Bell inequalities}

\author{A. Sauer}
\address{Institut f\"{u}r Angewandte Physik, Technische Universit\"{a}t
Darmstadt, D-64289 Darmstadt, Germany}
\ead{alexander.sauer@physik.tu-darmstadt.de}

\author{J. Z. Bern\'ad}
\address{Peter Gr\"unberg Institute (PGI-8), Forschungszentrum J\"ulich, D-52425
J\"ulich, Germany}
\address{Institut f\"{u}r Angewandte Physik, Technische Universit\"{a}t
Darmstadt, D-64289 Darmstadt, Germany}
\ead{j.bernad@fz-juelich.de}

\author{H. J. Moreno}
\address{Institut f\"{u}r Angewandte Physik, Technische Universit\"{a}t
Darmstadt, D-64289 Darmstadt, Germany}

\author{G. Alber}
\address{Institut f\"{u}r Angewandte Physik, Technische Universit\"{a}t
Darmstadt, D-64289 Darmstadt, Germany}

\date{\today}

\begin{abstract}
Euclidean volume ratios between quantum states with positive partial transpose and all quantum states in bipartite systems are investigated. 
These ratios allow a quantitative exploration of the typicality of entanglement and of its detectability by Bell inequalities.
For this purpose a new numerical approach is developed. 
It is based on the Peres-Horodecki criterion, on a characterization of the convex set of quantum states by inequalities resulting from Newton identities and from Descartes' rule of signs, and on a numerical 
approach involving the multiphase  Monte Carlo method and the hit-and-run algorithm. This approach confirms not only recent analytical and numerical results on two-qubit, qubit--qutrit, and qubit--four-level qudit states 
but also allows for a numerically reliable numerical treatment of so far unexplored qutrit--qutrit states. Based on this numerical approach with the help of the 
Clauser-Horne-Shimony-Holt inequality and the
Collins-Gisin inequality the degree of detectability of entanglement is investigated for two-qubit quantum states. 
It is investigated quantitatively to which extent a combined test of both Bell inequalities can increase the detectability of entanglement beyond what is achievable by each of these inequalities separately.
\end{abstract}
\vspace{2pc}

\submitto{\JPA}

\maketitle

\section{Introduction}

Entanglement is one of the characteristic quantum phenomena of distinguishable composite quantum systems \cite{Nielsen,Peres}. Therefore, questions concerning how to distinguish
entangled and separable quantum states and how to quantify their typicality play an important role in quantum information science \cite{Zyczkowski0}. 
For two special cases, namely for two-qubit and for qubit-qutrit states
a simple necessary and sufficient condition for identifying entanglement is known, the Peres-Horodecki criterion \cite{Peresp, Horodecki}. For these special quantum systems complications originating from bound 
entanglement do not arise and therefore all quantum states having positive partial transpose (PPT) are separable. Thus, in these quantum systems a convenient measure for 
the typicality of separability and thus also of entanglement is the relative volume of PPT quantum states in the space of all possible quantum states.  After the early work of Zyczkowski et al. \cite{Zyczkowski1, Zyczkowski2} 
numerous investigations have been performed aiming at estimating the volumes of separable and entangled states by various volume measures 
\cite{Zyczkowski3,Zyczkowski4,Slater1,Slater2,Slater3,Slater4,Shang,Seah,Milz,Lovas,Fei} and by
using various approaches, such as Bloore's representation \cite{Bloore}, Bures' metric \cite{Caves}, or the Ginibre ensemble \cite{GinibreHitRun} combined with Monte Carlo strategies \cite{Liu}.

Although by now numerous results are available estimating not only volumes of separable and entangled states but also volume ratios of PPT and non-PPT quantum states, there is still a need for new ideas capable of estimating these volume ratios accurately also for higher dimensional bipartite quantum systems. It is a main purpose of this paper to present and test such a new numerical approach by applying it to the estimation of the typicality of PPT quantum states and consequently also of non-PPT entangled quantum states. For this purpose a systematic 
approach is developed based on measuring these typicalities in several quantum systems by using an Euclidean volume measure. This is possible because the vector space of 
square matrices has a natural scalar product, namely the Hilbert-Schmidt inner product. Therefore, each square matrix can be considered as a point in an Euclidean vector space with a well defined 
associated volume measure. Furthermore, the possible quantum states are described by all possible positive semidefinite matrices with unit trace. It is known that the convex set formed by all 
quantum states can be described in a convenient way by inequalities resulting from an application of Newton identities \cite{Horn} and Descartes' rule of signs \cite{D1,D2} to characteristic 
polynomials of density matrices describing these quantum states \cite{Siennicki,Kimura,Byrd,Kryszewski,Gamel}. As a result the Euclidean volumes of the convex set of all possible quantum states 
and of the convex set of all PPT quantum states can be estimated numerically by a combination of the Muller \cite{Muller59,BoxM,Harman} and multiphase \cite{Kannan, Sim03} Monte Carlo methods
and of the hit-and-run algorithm \cite{Smith,LovaszV,LovaszV2}. 
The main advantage of the use of the Newton identities in this context lies in the reduced number of arithmetic operations required. This reduction of complexity is possible, because for deciding 
whether a given Hermitian $n\times n$ matrix is a quantum state or not, we need only a test for non-negativity of the eigenvalues and not their precise values.
Based on this approach it is demonstrated that not only known results on the typicality of entangled quantum states can be 
confirmed in a unified way, but also new reliable results on the typicality of PPT quantum states can be obtained in higher dimensional bipartite quantum systems.

Another purpose of this paper is to explore the detectability of bipartite entanglement by violations of 
Bell inequalities\cite{Pitowsky,Bell}. In particular, on the basis of our numerical Monte-Carlo approach we investigate the Euclidean volume ratio between entangled two-qubit quantum states violating 
Bell inequalities and all entangled states. For this purpose we concentrate on two types of Bell inequalities, the Clauser-Horne-Shimony-Holt (CHSH) \cite{CHSH} 
and the Collins-Gisin inequality \cite{CG}.
 Bell inequalities define half-spaces which are convex sets in the Euclidean vector space of the Hilbert-Schmidt inner product. 
These half-spaces contain the set of all separable quantum states. 
Thus, all Bell inequalities define a common non-empty convex set \cite{Rockafellar} which is larger than the convex set of all separable states. 
The quantum states belonging to this common convex set are not able to violate any kind of Bell inequality and are thus consistent with local realistic theories. Within this geometrical context, we are able to 
compare the performance of the CHSH and the Collins-Gisin type Bell inequalities with respect to detectability of entanglement. It is shown that each of these 
two types of Bell inequalities is capable of detecting only a 
small fraction of all entangled states. As there are entangled quantum states which violate only one of them but not the other one \cite{CG}, it is demonstrated that the combination of both types of 
inequalities is able to detect significanlty more entangled quantum states.

The paper is organized as follows. In Sec. \ref{I} the necessary and sufficient conditions are discussed under which a self-adjoint matrix
is positive semidefinite and describes a quantum state. With the help of Newton identities
and Descartes' rule of signs these conditions can be described systematically by a set of inequalities characterizing the convex set of quantum states. In Sec. \ref{II} the isomorphism between
the set of self-adjoint matrices with unit trace and points in an Euclidean vector space over the field of real numbers is used to develop two numerical Monte Carlo procedures.
These numerical procedures are based on the Muller method, the multiphase Monte Carlo method and the hit-and-run algorithm.
Numerical results for different classes of two-qubit and qubit-qutrit states are discussed in Sec. \ref{III} including also general qutrit-qutrit states and qubit-qudit states for four-level qudits.
In Sec. \ref{IV} with the help of the CHSH and the Collins-Gisin Bell inequalities the ratios between detectable entangled states and all entangled states are investigated for different classes of two-qubit states. A summary and
concluding remarks are presented in Sec. \ref{V}.

\section{Characterization of the convex set of quantum states in an Euclidean space}
\label{I}

In this section a general mathematical framework is presented for describing 
quantum states of an $n$ dimensional Hilbert space as elements of a convex set embedded in a $d=n^2-1$ dimensional real Euclidean vector space. 
Thereby the positive semidefiniteness of quantum states is taken into account by a set of inequalities which originate from applying Newton identities \cite{Horn} 
and Descartes' rules of signs \cite{D1,D2} to the characteristic polynomials of self-adjoint matrices with unit trace. The purpose of this section is to summarize the key ingredients of this approach, which can 
also be found in Refs. \cite{Kimura, Byrd, Kryszewski, Gamel}.

We consider the finite dimensional vector space $\mathbb{C}^n$ whose elements are represented in the canonical basis by all $n$-tuples of complex numbers. With the scalar product
\begin{equation}
 \langle x, y\rangle=\sum^n_{i=1} \overline{x}_i y_i=(\overline{x}_1,\overline{x}_2, \dots, \overline{x}_n) \left(
\begin{array}{c}
y_1\\
y_2\\
. \\
.\\
y_n
\end{array}
\right), \quad \forall x,y \in \mathbb{C}^n \nonumber
\end{equation}
this vector space is a Hilbert space. Thereby, $\overline{z}$ is the complex conjugate of the complex number $z \in \mathbb{C}$. This scalar product is antilinear in the first and linear in the second variable.
The norm $\langle x,x \rangle^{1/2}$ of any element $x$ in $\mathbb{C}^n$ will be denoted by $\|x\|$. The space of $n \times n$ matrices with complex entries $M_n(\mathbb{C})$ 
can be identified with the linear operators of this $n$ dimensional Hilbert space $\mathbb{C}^n$ if a canonical orthonormal basis is fixed. Keeping this identification in mind in the 
following we shall no longer distinguish between linear operators and their representations as matrices in $M_n(\mathbb{C})$.
The adjoint of a matrix $A$ is the unique matrix $A^\dagger$ satisfying $\langle A^\dagger x, y \rangle=\langle x, Ay \rangle$ for all $x,y$ in $\mathbb{C}^n$, or in other words 
the complex conjugate of the transpose of $A$. The trace $\mathrm{Tr}\{A\}$ of $A\in M_n(\mathbb{C})$ is given by the sum of its diagonal matrix elements and
is independent of the choice of an orthonormal basis. 

The $n^2$ dimensional vector space $M_n(\mathbb{C})$ together with the
Hilbert-Schmidt scalar product $\langle A,B\rangle_{\text{HS}}=\mathrm{Tr} \{A^\dagger B\}$ with $A,B \in M_n(\mathbb{C})$ constitutes an $n^2$ dimensional Hilbert space.  An elementary orthonormal basis in 
$M_n(\mathbb{C})$ with respect to this Hilbert-Schmidt scalar product is given by the $n \times n$ matrices $\{(E_{i,j})_{a,b}\}_{1 \leqslant i,j,a,b \leqslant n}$ with
$(E_{i,j})_{a,b}  = \delta_{ia}\delta_{jb}$ with $\delta_{ia},\delta_{jb}$ denoting Kronecker delta functions. A convenient orthonormal basis for the subspace of self-adjoint matrices can be constructed with 
the help of the $n^2-1$ traceless orthogonal self adjoint generators $T_i = T_i^{\dagger},~2\leq i \leq n^2$ of
 the Lie group $SU(n)$ which can be chosen such that they fulfill the orthonormality conditions\cite{Kimura}
\begin{eqnarray}
 \mathrm{Tr} \{T_iT_j\}= \delta_{ij}, \quad i=2,\dots, n^2. 
\end{eqnarray}
Together with the properly normalized unit matrix $T_1:= I_n/\sqrt{n}$ they form
a convenient orthonormal basis, which allows to identify every self-adjoint matrix $A \in M_n(\mathbb{C})$ by $n^2$ independent real-valued parameters $a_i \in \mathbb{R},1\leq i\leq n^2$ according 
to the relation \cite{Rudin}
\begin{eqnarray}
A &=&
\sum^{n^2}_{i=1} \langle T_i, A \rangle _{\text{HS}} T_i =
\sum_{i=1}^{n^2} \mathrm{Tr}\{T_i A\}T_i  \equiv \sum_{i=1}^{n^2}a_i T_i.
\label{repr}
\end{eqnarray}
In this basis the norm $||A||_{\text{HS}}$ of this self-adjoint operator $A$ is given by the relation
\begin{eqnarray}
 ||A||^2_{\text{HS}}&=&\langle A, A \rangle _{\text{HS}}=\sum^{n^2}_{i=1} \langle T_i, A \rangle^2.
\end{eqnarray}
For $n=2$ the normalized unit matrix  $I_2/\sqrt{2}$ together with the normalized Pauli matrices
$ \sigma_{i}/\sqrt{2}$ with $i=x,y,z$ are an example of such an orthonormal self-adjoint basis involving the generators of the Lie group $SU(2)$. For $n=3$ we have the normalized unit matrix $I_3/\sqrt{3}$ together with the eight normalized Gell-Mann matrices (see Eq.\eqref{Gell-Mann}).  
As an outlook, it is worth mentioning that there are also other interesting orthonormal bases \cite{Krammer}, 
though not all are suitable for our approach based on real vector spaces.

Within the subspace of self-adjoint matrices the set of density matrices describing quantum states is given by
the subset of positive semidefinite matrices with unit trace, i.e.
\begin{eqnarray}
 \mathcal{D}(\mathbb{C}^n)&=&\{\rho \in M_n(\mathbb{C}): \rho \geqslant 0, \quad \mathrm{Tr} \{\rho\}=1 \}.
\end{eqnarray}
Therefore, not every self-adjoint matrix of the form of Eq.(\ref{repr}) with unit trace, i.e. $\mathrm{Tr}\{\rho\} = 1$ or equivalently $a_1 = 1/\sqrt{n}$, is a quantum state. 
In order to characterize the positive semidefiniteness of quantum states in an efficient way we start from
the characteristic polynomial $p_A(\xi)$ of an arbitrary self-adjoint matrix $A \in M_n(\mathbb{C})$, i.e.,
\begin{eqnarray}
 p_A(\xi)&=&\det (\xi I_n-A)=\sum^n_{k=0} (-1)^k c^{(n)}_k \xi^{n-k} \label{chPolynom}
\end{eqnarray}
with
\begin{eqnarray}
 c^{(n)}_0 &=& 1,~
 c^{(n)}_1=\sum^n_{i=1} \lambda_i,~~
 c^{(n)}_2 = \sum_{1\leqslant i <j \leqslant n} \lambda_i\lambda_j,\cdots \nonumber\\
 c^{(n)}_n&=&\lambda_1\lambda_1 \dots \lambda_n
\end{eqnarray}
and with $\lambda_i\in {\mathbb R}$ ($1\leq i\leq n$) denoting the eigenvalues of $A$. All coefficients $c^{(n)}_i$ are elementary symmetric functions of these eigenvalues and can 
be related to traces of powers of the linear operator $A$ of the form
$p_k=\mathrm{Tr} \{A^k\}$ by the Newton identities \cite{Horn}, i.e.
\begin{eqnarray}
 p_1&=&c^{(n)}_1, \nonumber \\
 p_k&=&\sum^{k-1}_{i=1} (-1)^{i+1} c^{(n)}_i p_{k-i}+(-1)^{k+1}k c^{(n)}_k, \quad 1<k\leqslant n, \nonumber \\
 p_k&=&\sum^{k-1}_{i=1} (-1)^{i+1} c^{(n)}_i p_{k-i}, \quad k>n. \label{Newton0}
\end{eqnarray}
For a self-adjoint matrix with unit trace, i.e. $\mathrm{Tr} \{A\}=1$,
all coefficients $c^{(n)}_i$ can be obtained from these Newton identities recursively, i.e., 
\begin{eqnarray}
 &&c^{(n)}_1=1, \label{Newton} \\
 &&c^{(n)}_2=\frac{1}{2}-\frac{1}{2} p_2, \nonumber \\
 &&c^{(n)}_3=\frac{1}{6}-\frac{1}{2} p_2+\frac{1}{3}p_3, \nonumber \\
 &&c^{(n)}_4=\frac{1}{24}-\frac{1}{4} p_2+\frac{1}{3}p_3+\frac{1}{8} p^2_2-\frac{1}{4} p_4,
\dots \nonumber 
\end{eqnarray}
The characteristic polynomial $p_A(\xi)$ of a self-adjoint matrix $A$ has only real valued roots \cite{Rudin}. Therefore, according to Eq. \eqref{Newton} $p_A(\xi)$ is a 
polynomial with real valued coefficients so that
Descartes' rule of signs \cite{D1,D2} can be used to address the question of positive semidefiniteness of $A$. This rule states that the number of positive real roots of the polynomial
$p_A(\xi)$ equals the number of sign
changes in the sequence of coefficients. Therefore, based on Eq. \eqref{chPolynom} we have
\begin{equation}
 A\geqslant0 \Leftrightarrow c^{(n)}_k \geqslant 0, \quad \forall k \in {0,1,\dots,n} . \label{condition}
\end{equation}
It is evident from Eq. \eqref{Newton} that a self-adjoint matrix $A$ with unit trace is a quantum state, i.e. $A \in \mathcal{D}(\mathbb{C}^n)$, iff
\begin{eqnarray}
 &&\frac{1}{2}-\frac{1}{2} p_2\geqslant0, \label{Newtonuse} \\
 &&\frac{1}{6}-\frac{1}{2} p_2+\frac{1}{3}p_3\geqslant0, \nonumber \\
 &&\frac{1}{24}-\frac{1}{4} p_2+\frac{1}{3}p_3+\frac{1}{8} p^2_2-\frac{1}{4} p_4\geqslant0, \cdots\nonumber
\end{eqnarray}
These conditions fully characterize the positive semidefiniteness of self-adjoint matrices with unit trace. 
The Monte Carlo algorithms presented in the following sections start from the representation of the convex set of quantum states
of an $n$ dimensional Hilbert space by Eq.(\ref{repr}) and by the inequalities (\ref{Newtonuse}).

\section{Numerical Monte Carlo methods}
\label{II}

In this section we describe two numerical methods for the estimation of volume ratios of convex bodies. The primary task is 
to compute numerically the volumes of convex sets of the form
\begin{equation}
 V_C=\int_{M_n(\mathbb{C})}d \mu~ \chi_K  \label{integral}
\end{equation}
where $\chi_K$ is the characteristic function of the convex set $K$ of interest and $\mu$ is a volume measure on $M_n(\mathbb{C})$. 
As shown
in Sec. \ref{I} the set of self-adjoint matrices forms a subspace
in $M_n(\mathbb{C})$ isomorphic to $\mathbb{R}^{n^2}$. Therefore, the Hilbert-Schmidt norm of the difference between
two self-adjoint matrices, say $A$ and $B$, is connected to the Euclidean distance function in $\mathbb{R}^{n^2}$ by the relation
\begin{eqnarray}
 \|A-B\|_{{\text{HS}}}&=&\left\|\sum_{i=1}^{n^2} (a_i - b_i) T_i\right\|_{{\text{HS}}} = \sqrt{\sum_{i=1}^{n^2} (a_i-b_i)^2}\nonumber
 \\
\end{eqnarray}
with $a_i,b_i \in \mathbb{R}$ and with $T_i$ ($1\leq i\leq n^2$) denoting an orthonormal basis of self-adjoint $n\times n$ matrices according to Eq. \eqref{repr}.
This implies that the measure $\mu$ in Eq. \eqref{integral}
is the volume of measurable subsets of the $n^2$ dimensional Euclidean space. 

A key ingredient of volume estimates is to generate random points uniformly over the corresponding convex set. However, the dimension of the Euclidean space containing this convex body plays an 
important role in the efficiency of the random point generator. In the following we investigate an exact random point generation and an approximate generation method, 
or Markov chain Monte Carlo sampler. Both approaches belong to the acceptance-rejection method and they have their pros and cons. The exact random point generator guarantees that we sample from a uniform 
distribution on a $d$ dimensional ball but not all the points will lie inside the convex body in question, which has a non-empty intersection with this $d$ dimensional ball.
The Markov chain Monte Carlo sampler 
generates points from the convex body, but the convergence to a uniform distribution requires more and more points as we increase the dimension of the Euclidean space.

\subsection{Multiphase Monte Carlo method}

The first numerical approach employs the Muller method \cite{Muller59, BoxM,Harman}, which is applied to generate random points uniformly in the $d$ dimensional ball, 
and combines it with the multiphase Monte Carlo method by using a 'sandwiching' technique \cite{Kannan, Sim03}. For this purpose $d$ dimensional vectors $\vec{v}$ are drawn from the uncorrelated 
multivariate normal distribution.  If, in addition, the random variable $u$ is distributed uniformly in the unit interval $[0,1]$, 
the vectors $ \vec{u}=r\,u^{1/d}\, \vec{v}/\sqrt{\vec{v} \cdot \vec{v}}$ are randomly and uniformly 
distributed in the $d$ dimensional ball with radius $r$. Furthermore, we consider $m$ concentric 
$d$ dimensional balls with radii 
$r_1<r_2<\dots<r_m$ around the origin of the convex set of interest, say $K$, and we apply the Muller method within each of these balls. According to Ref. 
\cite{Sim03} the natural number $m$ should be larger than $d \log d$. The Euclidean volume of the convex set of interest $K$ can be estimated by the 'product estimator'
\begin{eqnarray}
\mbox{vol}(K) &=& \mbox{vol}(K_1)\,\displaystyle\prod_{i=2}^m\frac{\mbox{vol}(K_i)}{\mbox{vol}(K_{i-1})}
\end{eqnarray}
with
 $K_i = K\cap \mathbb{B}(0, r_i) $
denoting the intersection between the convex set $K$ and the $d$ dimensional ball $\mathbb{B}(0, r_i)$ with radius $r_i$ and center at the origin of the Euclidean space. 
It is apparent that $K_1\subseteq K_2 \subseteq \dots \subseteq K_m$. 
Each domain $K_i$ with $i=1,\cdots,m$ or 'phase'
requires the generation of uniformly distributed independent points for estimating its Euclidean volume $\mbox{vol}(K_i)$.  However, for high dimensional convex sets this algorithm may already break down before the last 'phase' is reached, because the number of states found is too small for a satisfactory statistics.
With each new ball in the sequence more and more points need to be generated which requires increasing running times of this algorithm. 
Therefore, eventually the application of this algorithm is limited by current capacities  of nowadays computers. 
Our implementation of this algorithm involves manageable steps and stops whenever the number of points found inside $K$ is too small. We set this threshold number to be $10$.

\subsection{Hit-and-run algorithm}

Our second numerical approach has been introduced by Smith \cite{Smith} to generate points uniformly distributed within an arbitrarily bounded region. Thus, it is applicable to a convex body $K$.
This sampler makes a transition from a point $a \in K$ to another point $a' \in K$ by generating a direction vector $\vec{x}$ uniformly on the surface of a $d$ dimensional unit ball 
with center $a$ followed by generating a point $a'$ uniformly distributed on the line segment created by the intersection of $K$ and the line through $a$ with direction $\vec{x}$. 
The unit vector $\vec{x}$ is generated with the 
help of the Muller method and $a'$ is chosen by employing a one dimensional acceptance-rejection method on the line segment, i.e., we accept this point only if it lies in $K$. Now, we set $a'$ to be our starting 
point and we repeat the procedure. This algorithm realizes a random walk inside $K$ 
that converges efficiently to a uniform distribution and this is independent of the starting point inside $K$ \cite{LovaszV}. 
It is worth noting that one can combine the multiphase Monte Carlo technique with the hit-and-run algorithm in order to obtain even more efficient volume 
estimates \cite{LovaszV2}, but this is left out by us for future investigations.

\section{Euclidean volume ratios for bipartite quantum states with positive partial transpose}

\label{III}

In this section
we investigate numerically the Euclidean volume ratios $R$ between bipartite quantum states with positive partial transpose (PPT) and all bipartite quantum states. 
In two-qubit and qubit-qutrit systems the Euclidean volume ratio $R$ determines the volume 
ratio $R/(1-R)$ between separable and entangled states, because in these systems all quantum states with negative partial transpose are entangled. 
Our aim is to provide new estimates of $R$ for several bipartite quantum systems and to assess the reliability and efficiency of 
our numerical approaches by comparing our estimates with known analytical and numerical results.

In general, the set $\mathcal{D}(\mathbb{C}^n)$ of all quantum states is a convex set because
any convex combination of two density matrices is also a
density matrix. A density matrix $\rho_{AB}$ of a bipartite system with constituents $A$ and $B$ is called separable if it can be written as a convex combination of product states, i.e.
\begin{eqnarray}
 \rho_{AB}&=&\sum_k p_k \rho^{(A)}_k \otimes \rho^{(B)}_k, \quad 0\leqslant p_k \leqslant1,\quad \sum_k p_k=1,
 \nonumber\\
\end{eqnarray}
where $\rho^{(A)}_k$ ($\rho^{(B)}_k$) is a possible quantum state of system $A$ ($B$). It is clear from this definition of separability that the set of separable quantum states also forms a convex set.

Let us consider a finite dimensional bipartite quantum system with Hilbert space $\mathbb{C}^{n_A}\otimes\mathbb{C}^{n_B}$, where $n_A$ and $n_B$ are the dimensions of the subsystems. The map $\rho 
\rightarrow (\mathrm{\tau}_{n_A}\otimes\mathbb{I}_{n_B})\rho$ with the identity operation $\mathbb{I}_{n_B}$ on $M_{n_B}(\mathbb{C})$ is called partial transposition and is defined with respect to the 
canonical product basis as $\bra{ij} (\mathrm{\tau}_{n_A}\otimes\mathbb{I}_{n_B})\rho \ket{kl}=\bra{kj} \rho \ket{il}$. If a state is separable its density matrix has a positive partial transpose (PPT), i.e,
the result of the map is again a density matrix. All states having positive partial transpose are called PPT quantum states. They form a convex set. This procedure is independent of the subsystem that is 
transposed, because the eigenvalues of a square matrix are 
equal to the eigenvalues of its full transpose. For example, the transposition operator $\mathrm{\tau}_2$, which acts on qubits, has the following properties
\begin{eqnarray}
  \mathrm{\tau}_2\, I_2&=I_2, \quad \mathrm{\tau}_2 \sigma_x&=\sigma_x, \nonumber \\
  \quad \mathrm{\tau}_2 \sigma_y&=-\sigma_y, \quad \mathrm{\tau}_2 \sigma_z&=\sigma_z \label{transpose}
\end{eqnarray}
with $\{\sigma_x, \sigma_y, \sigma_z \}$ denoting the Pauli spin matrices. According to the Peres-Horodecki criterion \cite{Peresp,Horodecki} in two special cases, 
namely for two-qubit systems with Hilbert space
$\mathbb{C}^2\otimes\mathbb{C}^2 \cong \mathbb{C}^4$ and for qubit-qutrit systems with Hilbert space
$\mathbb{C}^2\otimes\mathbb{C}^3\cong \mathbb{C}^6$, all PPT quantum states are separable, i.e., the so-called phenomenon of bound entanglement or entangled PPT quantum states does not occur in these cases.

Let us first of all discuss  our numerical approach for estimating Euclidean volume ratios $R$ between bipartite PPT quantum states and all bipartite quantum states with the help of the multiphase Monte Carlo method. As quantum states have unit trace in the following we restrict ourselves to the subspace of 
self-adjoint matrices $A$ with unit trace, i.e. $a_1=1/\sqrt{n}$. According to Eq. \eqref{repr} an arbitrary 
element $A$ of this $d=(n^2-1)$ dimensional subspace is identified by its real-valued coordinates $(a_2,\dots, a_{n^2}) \in \mathbb{R}^{n^2-1}$. For a numerical estimate of the Euclidean volume 
ratio $R$ it is necessary to generate vectors of this type randomly and uniformly at first. For this purpose it is convenient to take also into account the first constraint of \eqref{Newtonuse} 
as a necessary condition for quantum states, i.e.
\begin{eqnarray}
 1 \geqslant \mathrm{Tr}\{A^2\}& \Leftrightarrow& \frac{n-1}{n} \geqslant \sum^{n^2}_{i=2} a^2_i, \label{lball}
\end{eqnarray}
so that the point $(a_2,\dots a_{n^2})$ is element of the $d=(n^2-1)$ dimensional ball with radius $r_n=\sqrt{n-1}/\sqrt{n}$. Therefore, the starting point
of the multiphase Monte Carlo method is that we choose 
$r_m\equiv r_n=\sqrt{n-1}/\sqrt{n}$ to be the radius of the largest ball containing all quantum states. The smallest radius $r_1$ can be chosen in a convenient way with the help of
Mehta's lemma \cite{Mehta89}. This lemma states that a self-adjoint matrix $A$ in $\mathbb{C}^n$ is positive if
\begin{eqnarray}
\mathrm{Tr}\{A^2\}&\leqslant& \frac{1}{n-1}.
\end{eqnarray}
Taking into account that the partial transposition required for an application of the Peres-Horodecki criterion leaves the Hilbert-Schmidt norm invariant, it is apparent that the $d$ 
dimensional ball with radius $1/\sqrt{n(n-1)}$ is a subset of
the separable states \cite{Zyczkowski0, Mehta89}. Therefore, it is convenient to choose 
$r_1=1/\sqrt{n(n-1)}$ for the radius of
the smallest $d$ dimensional ball. In order to select quantum states randomly, 
in a second step for each of these sampled points, say $A$, in each of the $d$ dimensional balls the remaining constraints in \eqref{Newtonuse} have to be tested in order to determine whether the randomly selected 
matrix $A$ belongs to the convex set of quantum states or not.  Due to Mehta's lemma $R_1=1$ and the ratio $R$ between the volumes of separable states and all states is estimated by
\begin{equation}
R = R_1\,\displaystyle\prod_{i=2}^m\frac{R_i}{R_{i-1}}. \label{ratio}
\end{equation}
It is clear from Eq. \eqref{ratio} that if we can obtain $R_m$ then $R=R_m$. However, generating uniformly random points in the largest balls in high dimensional spaces with the Muller method is subject
to the so-called "curse of dimensionality" phenomenon \cite{curseofdimensionality}.

In order to obtain reliable statistics we repeat this numerical procedure
several times. Denoting the number of these sampling repetitions by $s$
we eventually obtain an arithmetic mean ratio
\begin{eqnarray}
 \bar{R}&=&\frac{\sum^s_{k=1} R_k}{s} 
\end{eqnarray}
based on the individual results  $R_k$ ($k=1, \dots, s$) of these repetitions. This is the main estimation parameter for the Euclidean volume ratio between PPT quantum states and all quantum states. 
The standard deviation of the sample is then given by
\begin{equation}
 \sigma=\sqrt{\frac{1}{s-1} \sum^s_{k=1} \left(R_k-\bar{R}\right)^2}.
\end{equation}

In the case of the hit-and-run algorithm we are allowed to choose the starting point arbitrarily. Therefore, we start with $(a_2,\dots a_{n^2})=(0,0,\dots 0)$, i.e., the origin of the $d$ dimensional 
Euclidean space or the maximally mixed state. We apply the acceptance-rejection method to the next point by testing for the constraints of \eqref{Newtonuse}. The only difficulty is that the boundaries of the 
line segments are hard to determine.
Therefore we approximate the boundary in each direction by checking whether the point with a distance of $b_0=\sqrt{n-1}/\sqrt{n}$ to the starting point fulfills the constraints. 
If it does, which is rarely true, the point with a distance of $2b_0$ to the starting point is used as the upper bound.
If not, the procedure is repeated for $b_{i+i}=b_i/2$ until the constraints are fulfilled by $b_{i+i}$ such that $b_i$ can be set as the upper bound.
As we know that the $d$ dimensional ball with radius $r_n=\sqrt{n-1}/\sqrt{n}$ is a closing convex body for the set of all quantum states, this method will always yield an upper bound on the boundary of the line 
segment. 
Furthermore, at least half of the resulting line segment intersects with the set of all quantum states.
Then, random points are sampled from this line segment until the chosen point fulfills the constraints of \eqref{Newtonuse}.
This is used as the starting point for the next iteration. 
To obtain the standard deviation of the sampling, all obtained points are grouped into blocks of size $N_B$. This procedure for
obtaining the number of points $N_{\text{B}_i,\text{PPT}}$  fulfilling the PPT criterion within each block can be viewed as a Bernoulli trial with a success probability of $R$, 
if all points are independent samples 
from the set of all quantum states.
For large block sizes, the distribution of $N_{B_i, PPT}$ can then be approximated by a Gaussian with mean $R \cdot N_B$ and a standard deviation $\sigma_B$ which depends on the block size and on the independence 
of the points sampled by the 
hit-and-run algorithm. If not stated otherwise, a block size of $10^6$ points was used.
As we are interested in the standard deviation of the mean $\sigma_{\bar{R}}$, the number of blocks $N_I$ is taken into account to get $\sigma_{\bar{R}}^2 = \sigma_B^2/N_I $.

\subsection{Two-qubit Bell diagonal states}\label{sec:bell-dia}

\begin{figure}
   \centering
  \includegraphics[width=0.4\textwidth]{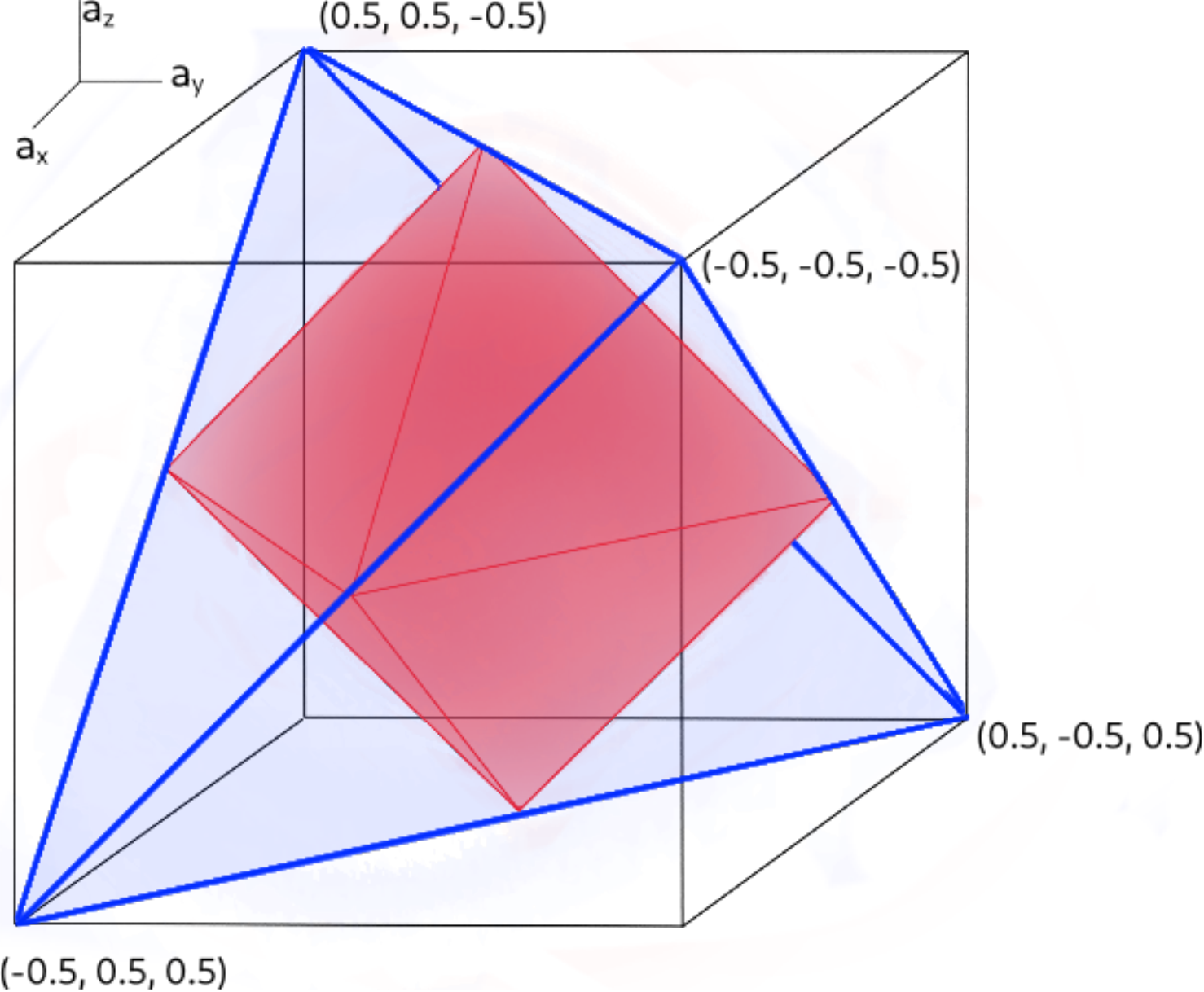}
     \caption{Schematic representation of the convex set of separable Bell diagonal two-qubit states (octahedron) inside  the convex set of all Bell diagonal two-qubit states (tetrahedron): 
     Both convex sets are embedded in the $d=3$ dimensional Euclidean space of all self-adjoint linear operators of the form of Eq. \eqref{eq:belldiag}.The four vertices of the tetrahedron represent the four maximally entangled two-qubit Bell states.}
   \label{Belldiagonal}
 \end{figure}

As a first example we consider Bell diagonal two-qubit states for which the Euclidean volume ratio $R$ is known analytically.
Bell diagonal two-qubit states are characterized by three real-valued independent parameters. They form a $d=3$ dimensional convex set embedded in the $d=3$ dimensional linear subspace of self-adjoint 
matrices with unit trace, and their representation reads
\begin{equation}
    \rho = \frac{I_4}{4}+\frac{1}{2}\sum_{i=x,y,z}\,a_i\,\sigma^{(A)}_i\otimes
    \sigma^{(B)}_i 
    \label{eq:belldiag}
\end{equation}
with $a_i \in \mathbb{R}$.
$A$ and $B$ label the two distinguishable subsystems and $\{\sigma_x, \sigma_y, \sigma_z \}$ are the Pauli spin matrices. 
Newton identities and the related inequalities \eqref{Newtonuse} determine the possible values of the three parameters $a_i$, which restrict self-adjoint operators of the form of Eq.\eqref{eq:belldiag} 
to quantum states. It has been shown (cf. Fig.\ref{Belldiagonal}) that the state space of Bell diagonal two-qubit states is a tetrahedron with separable quantum states forming an octahedron inside this tetrahedron  \cite{Ziman, LeMyOv06}. Therefore, the Euclidean volume ratio of Bell diagonal two-qubit states can be determined analytically. It is given by $R=0.5$.
For this case the multiphase Monte Carlo (MMC) method with $s=100$ repetitions and $10^8$ points generated in each $3$ dimensional ball yields the numerical result
\begin{equation}
 \bar{R}^{MMC}=0.4999, \quad \sigma^{MMC}=0.0001. 
\end{equation}
The analytical value of $R = 0.5$ is also confirmed by the hit-and-run (HR) algorithm with $5.4 \times 10^9$ quantum states and the numerical estimate is
\begin{equation}
 \bar{R}^{HR}=0.499998, \quad \sigma^{HR}=0.000014. 
\end{equation}

\subsection{Two-qubit X-states}\label{sec:xsta}

Two-qubit X-states represent another class of quantum states which has received considerable attention for purposes of quantum information processing \cite{Ravi09}. These states are characterized by 
seven independent real-valued parameters, i.e., $d=7$, and have the form
\begin{equation}
 \rho=\begin{pmatrix}
\rho_{11} & 0 & 0 & \rho_{14} \\
0 & \rho_{22} & \rho_{23} & 0 \\
0 & \rho_{32} & \rho_{33} & 0 \\
\rho_{41} & 0 & 0 & \rho_{44}
\end{pmatrix}. \nonumber
\end{equation}
Thus their definition is basis dependent.
Bell diagonal states are a subset of X-states. The Euclidean volume ratio between separable X-states and 
all X-states is analytically known to be given by $R=0.4$ \cite{Milz}. These X-states form a $7$ dimensional convex set within the $7$ dimensional subspace of self-adjoint matrices with unit 
trace, and their representation reads
\begin{equation}
\rho = \frac{I_4}{4} + \frac{1}{2}\sum_{i=1}^7 a_i\,T_i
\label{Xs}
\end{equation}
with $a_i\in \mathbb{R}$ and with
\begin{eqnarray}
T_1&=&\sigma^{(A)}_z\otimes I^{(B)}_2, \, T_2=I^{(A)}_2 \otimes \sigma^{(B)}_z, \, T_3=\sigma^{(A)}_x\otimes \sigma^{(B)}_x, \nonumber \\
T_4&=&\sigma^{(A)}_x\otimes \sigma^{(B)}_y, \, T_5=\sigma^{(A)}_y\otimes \sigma^{(B)}_x, \, T_6=\sigma^{(A)}_y \otimes \sigma^{(B)}_y, \nonumber \\
T_7&=&\sigma^{(A)}_z\otimes \sigma^{(B)}_z. \nonumber
\end{eqnarray}
The multiphase Monte Carlo method with $s=150$ repetitions and $10^7$ points generated in each $7$ dimensional ball yields the numerical result
\begin{equation}
 \bar{R}^{MMC}=0.3998, \quad \sigma^{MMC}=0.0005,
\end{equation}
and the hit-and-run algorithm with $4 \times 10^9$ quantum states results in
\begin{equation}
 \bar{R}^{HR}=0.400003, \quad \sigma^{HR}=0.000022. 
\end{equation}
Both numerical estimates are in very good agreement with the analytically known result of $R=0.4$.
Comparing this result with the result for Bell diagonal two-qubit states it is apparent that increasing the number $d$ of independent coefficients characterizing the two-qubit states
reduces the volume of the separable states inside the convex set of all two-qubit X-states. 

\subsection{Rebit-rebit states} \label{sec:rebqb}

Another interesting class of quantum states which has an analytical ratio of $R=29/64$ are the real valued two-qubit states \cite{Lovas}. They form a $9$ dimensional convex set within 
the $9$ dimensional subspace of self-adjoint matrices with unit trace. Using the notation of  \eqref{repr} they can be represented in the form
\begin{equation}
\rho = \frac{I_4}{4} + \frac{1}{2}\sum_{i=1}^9 a_i\,T_i
\label{Xs}
\end{equation}
with $a_i\in \mathbb{R}$. 
Thereby, the $9$ dimensional basis with elements $T_i$ with $1\leq i \leq 9$ has the be chosen in such a way that only real-valued basis vectors are included from the complete $15$ dimensional basis, namely
\begin{eqnarray}
T_1&=&I^{(A)}_2 \otimes \sigma^{(B)}_x, \, T_2=I^{(A)}_2 \otimes \sigma^{(B)}_z, \, T_3=\sigma^{(A)}_x\otimes I^{(B)}_2, \nonumber \\
T_4&=&\sigma^{(A)}_z\otimes I^{(B)}_2, \, T_5=\sigma^{(A)}_x\otimes \sigma^{(B)}_x, \, T_6=\sigma^{(A)}_x \otimes \sigma^{(B)}_z, \nonumber \\
T_7&=&\sigma^{(A)}_y\otimes \sigma^{(B)}_y, \, T_8=\sigma^{(A)}_z \otimes \sigma^{(B)}_x, \, T_9=\sigma^{(A)}_z \otimes \sigma^{(B)}_z. \nonumber
\end{eqnarray}
  
The multiphase Monte Carlo method with $s=200$ repetitions and $10^8$ points generated in each $9$ dimensional ball yields the numerical result
\begin{equation}
 \bar{R}^{MMC}=0.45309, \quad \sigma^{MMC}=0.0013 
\end{equation}
and the hit-and-run algorithm and $4 \times 10^9$ quantum states yields the numerical result
\begin{equation}
 \bar{R}^{HR}=0.453111, \quad \sigma^{HR}=0.000027. 
\end{equation}
These estimated values are very close to the analytical value of $R=29/64=0.453125$ \cite{Lovas}. This shows that our numerical approaches are capable of yielding  very accurate estimates also for this benchmark value. 

\subsection{General two-qubit states}\label{sec:tqb}

A general density matrix can be written in the form
\begin{eqnarray}
\rho &=& \frac{I_4}{4} +\frac{1}{2} 
 \sum_{i=x,y,z} \tau^{(A)}_i \,\sigma^{(A)}_i  \otimes I^{(B)}_2
+ \frac{1}{2} \sum_{i=x,y,z} \tau^{(B)}_i \,I^{(A)}_2 \otimes \sigma^{(B)}_i \nonumber\\
&+&
\frac{1}{2} \sum_{i,j=x,y,z} \nu_{i,j} \,\sigma^{(A)}_i  \otimes  \sigma^{(B)}_j
\label{full}
\end{eqnarray}
with $\tau^{(A)}_i,\tau^{(B)}_i,\nu_{i,j}\in \mathbb{R}$.
For these general two-qubit states the multiphase Monte Carlo method yields the numerical result 
\begin{equation}
 \bar{R}^{MMC}=0.243, \quad \sigma^{MMC}=0.007
\end{equation}
for a sample size of $10^8$ points in each $15$ dimensional ball and for the sampling repetition $s=150$. Alternatively the hit-and-run algorithm with $4 \times 10^9$ quantum states yields the numerical result
\begin{equation}
 \bar{R}^{HR}=0.242444, \quad \sigma^{HR}=0.000027. 
\end{equation}
These ratios are close to the recent combined analytical and numerical results of Slater and Dunkl \cite{Slater3} supporting the conjecture that $R=8/33\approx 0,24242$, and are consistent with the numerical result of 
Shang et al. \cite{Shang}, i.e. $R=0.242 \pm 0.002$, and of Milz et al. \cite{Milz}, i.e., $R=0.24262 \pm 0.0134$. Furthermore, our method is able to provide a very accurate estimate for a significantly smaller number of states than required in the recent
numerical study of Fei et al. \cite{Fei}, where $5 \times 10^{11}$ points have had to been sampled to obtain the value $R = 0.24243 \pm 0.00001$.

It is worth mentioning that the number of $10^8$ randomly generated points in the case of the multiphase Monte Carlo method becomes slightly problematic for the largest balls, because a few tens of states are found only.
This significant increase of the number of randomly selected points required can be made plausible by a simple qualitative argument. For this purpose let us consider the $d=n^2-1$ dimensional 
convex set of quantum states $K_d$ which is in contact with the largest $d$ dimensional ball $\mathbb{B}(0,r_n)$ of radius $r_n =\sqrt{n-1}/\sqrt{n}$  within which points have to be selected uniformly 
and randomly according to the Muller method or according to the last step of the multiphase Monte Carlo method. In $d$ dimensions their volume ratio is given by
$V_{d} = vol(K_d)/vol(\mathbb{B}(0,r_n))=q^d$ with $0 < q < 1$ and $q$  being a slowly varying function of $d$  \cite{Sim03}. This quantity measures the probability of finding  a point inside $K_d$. Therefore, 
finding with certainty a point inside $K_d$, i.e. a quantum state, requires at least the random selection of $N_d = q^{-d}$ uniformly distributed points. As a result of this scaling and under the 
simplifying assumption of a $d$ independent value of $q$, finding a quantum state
inside $K_{d'}$ in $d'$ dimensions  requires at least the random selection of $N_{d'} = q^{-d'} = N_d (N_d)^{(d'-d)/d}$
uniformly distributed points. This exponential increase of $N_{d'}$ with increasing dimensions $d'$ and the corresponding numerical problems also affected our numerical simulations already for $d=15$.
In our numerical simulations the number of quantum states found decreased significantly when
changing $d$ from $d=7$ to $d=15$. Thus, for $N=10^6$ randomly selected points, for example, the number of quantum states in the largest $d$ dimensional ball $\mathbb{B}(0,r_n)$ decreased from $3340$ 
in the case of X-states ($d=7$) to $34$ in the general two-qubit case ($d=15$).  From this observation one extrapolates that at 
least $N_{35} = 10^8\times (10^8)^{(35-15)/15} > 10^{18}$ points have to be selected randomly in the last step of the multiphase Monte Carlo method for finding a quantum 
state of a general qubit-qutrit system characterized by $n=6$ and $d=n^2 - 1 = 35$. In view of this considerable computational effort in the following the properties of qubit-qutrit states are explored 
only for a few subcases with the help of the multiphase Monte Carlo method and its efficiency is compared to the hit-and-run algorithm.

\subsection{A few families of qubit-qutrit states}
\label{sec:qbqtspec}

In this section numerical results are presented for the Euclidean volume ratios $R$ for some special cases of qubit-qutrit states.
In particular, results are presented for convex subsets of qubit-qutrit states which are embedded in linear subspaces of dimensions $d=8,12$ and $d=24$. 
In the case of a qubit-qutrit system a general density matrix can be written in the form
\begin{eqnarray}
\rho &=& \frac{I_6}{6} +  \frac{1}{\sqrt{6}}\sum_{i=x,y,z} \tau^{(A)}_i \,\sigma_i^{(A)} \otimes I^{(B)}_3 + \frac{1}{2}\sum_{j=1}^8 \tau^{(B)}_j \,I^{(A)}_2 \otimes \gamma^{(B)}_j 
  \nonumber\\
 &+&
 \frac{1}{2}\sum_{i=x,y,z} \sum_{j=1}^{8} \nu_{i,j} \,\sigma_i^{(A)} \otimes \gamma^{(B)}_j
\label{qubitqutrit}
\end{eqnarray}
with $\tau^{(A)}_i,\tau^{(B)}_j,\nu_{i,j}\in \mathbb{R}$ and the Gell-Mann matrices $\gamma_i$
\begin{eqnarray}
 \gamma_1 &=& \begin{pmatrix} 0 & 1 & 0 \\ 1 & 0 & 0 \\ 0 & 0 & 0 \end{pmatrix}, \, \gamma_2 = \begin{pmatrix} 0 & -i & 0 \\ i & 0 & 0 \\ 0 & 0 & 0 \end{pmatrix}, \,
 \gamma_3 = \begin{pmatrix} 1 & 0 & 0 \\ 0 & -1 & 0 \\ 0 & 0 & 0 \end{pmatrix}, \nonumber \\
\gamma_4 &=& \begin{pmatrix} 0 & 0 & 1 \\ 0 & 0 & 0 \\ 1 & 0 & 0 \end{pmatrix}, \, \gamma_5 = \begin{pmatrix} 0 & 0 & -i \\ 0 & 0 & 0 \\ i & 0 & 0 \end{pmatrix}, \,
\gamma_6=\begin{pmatrix} 0 & 0 & 0 \\ 0 & 0 & 1 \\ 0 & 1 & 0 \end{pmatrix}, \nonumber \\
\gamma_7 &=& \begin{pmatrix} 0 & 0 & 0 \\ 0 & 0 & -i \\ 0 & i & 0 \end{pmatrix}, \, \gamma_8= \frac{1}{\sqrt{3}} \begin{pmatrix} 1 & 0 & 0 \\ 0 & 1 & 0 \\ 0 & 0 & -2 \end{pmatrix}.
\label{Gell-Mann}
\end{eqnarray}
The qubit and qutrit subsystems are denoted by $A$ and $B$, respectively.

Let us consider three subspaces of increasing dimensions $d=8, 12$ and $d= 24$ which involve self-adjoint matrices $\rho$ of the form
\begin{eqnarray}
 (i) \ \ \quad \rho &=&\frac{I_6}{6}+\frac{1}{2}\sum_{i=1}^8 \nu_{y,i} \sigma^{(A)}_y \otimes \gamma^{(B)}_i \label{eq:qbqt-d8}, \\
 (ii) \  \quad \rho &=&\frac{{I_6}}{6}+\frac{1}{2}\sum_{i=1}^4 \nu_{x,i}\,\sigma^{(A)}_x \otimes \gamma^{(B)}_i \label{eq:qbqt-d12} \\
       &+& \frac{1}{2}\sum_{i=1}^4 \nu_{y,i}\,\sigma^{(A)}_y \otimes \gamma^{(B)}_i + \frac{1}{2}\sum_{i=1}^{4} \nu_{z,i}\,\sigma^{(A)}_z \otimes \gamma^{(B)}_i, \nonumber \\
 (iii) \quad \rho &=&\frac{{I_6}}{6}+\frac{1}{2}\sum_{i=1}^8 \nu_{x,i}\,\sigma^{(A)}_x \otimes \gamma^{(B)}_i  \label{eq:qbqt-d24} \\
    &+& \frac{1}{2}\sum_{i=1}^8 \nu_{y,i}\,\sigma^{(A)}_y \otimes \gamma^{(B)}_i + \frac{1}{2}\sum_{i=1}^{8} \nu_{z,i}\,\sigma^{(A)}_z \otimes \gamma^{(B)}_i. \nonumber 
\end{eqnarray}

Case $(i)$ with $d=8$ is an 
interesting special case. All quantum states within this $8$ dimensional subspace are separable so that we obtain the ratio $R=1$. 
This is due to the fact that
there 
is a local unitary transformation acting on qubit $A$, i.e.
\begin{equation}
 U^{(A)}=\begin{pmatrix}
          e^{-\frac{i \pi }{4}} \cos \left(\frac{\beta}{2}\right) & -e^{-\frac{i \pi }{4}} \sin \left(\frac{\beta}{2}\right) \\
 e^{\frac{i \pi }{4}} \sin \left(\frac{\beta}{2}\right) & e^{\frac{i \pi }{4}} \cos \left(\frac{\beta}{2}\right) \\
         \end{pmatrix}, \nonumber
\end{equation}
with $\beta \in [0,2 \pi]$. This unitary transformation has the characteristic property
\begin{equation}
 \left(U^{(A)}\right)^\dagger \sigma^{(A)}_y U^{(A)}=\cos \beta \sigma^{(A)}_x + \sin \beta \sigma^{(A)}_z. \nonumber
\end{equation}
As a local unitary transformation does not change the PPT property it is apparent from the transformation properties of the Pauli matrices under transposition that all states of the form of \eqref{eq:qbqt-d8} are PPT quantum states.

Numerical results for Euclidean volume ratios $R$  of the convex sets of quantum states within these linear subspaces are presented in Tables \ref{tab:qbqt} and \ref{tab:qbqtHR}.
\begin{table}[H]
  \centering
\begin{tabular}{ |c|c|c|c| }
\hline
Case & $N$ &  $\bar{R}^{MMC}$  & $\sigma^{MMC}$ \\
\hline
$(i)$ & $10^7$ &  $1.0$ & $0.0$\\
$(ii)$  &$10^7$ & $0.198$ & $0.029$ \\
$(iii)$ & $10^7$ & $0.016$  & $0.006$ \\
\hline
\end{tabular}
\caption{
Estimates of  the Euclidean volume ratios $R$ and their standard deviations $\sigma$ for three classes of qubit-qutrit states using the multiphase Monte Carlo method: 
$(i)$ states of the form of Eq. \eqref{eq:qbqt-d8} ($d=8$);
$(ii)$ states of the form of Eq.\eqref{eq:qbqt-d12} ($d=12$); $(iii)$ states of the form of Eq. \eqref{eq:qbqt-d24} ($d=24$). The sample size in each $d$ dimensional ball is denoted by $N$. The sampling repetition 
has been set to $s=50$.}
\label{tab:qbqt}
\end{table}
\begin{table}[H]
  \centering
\begin{tabular}{ |c|c|c|c| }
\hline
Case & $N$ &  $\bar{R}^{HR}$  & $\sigma^{HR}$ \\
\hline
$(i)$ & $10^7$ &  $1.0$ & $0.0$\\
$(ii)$  &$2\times 10^7$ & $0.1937$ & $0.0003$ \\
$(iii)$ & $7 \times 10^7$ & $0.02229$  & $0.00006$ \\
\hline
\end{tabular}
\caption{
Estimates of the Euclidean volume ratios $R$ and their standard deviations $\sigma$ for three classes of qubit-qutrit states using the hit-and-run algorithm: 
The three cases are the same as inTable \ref{tab:qbqt}. $N$ denotes the number of quantum states.}
\label{tab:qbqtHR}
\end{table}
In case $(ii)$ with $d=12$ there is a good agreement between the two Tables. However, consistent with the discussion of the previous subsection, the multiphase Monte Carlo technique with the Muller method 
has problems in finding enough quantum states for case $(iii)$ with $d=24$. This is the reason why the ratios are different for case $(iii)$ 
and it is also a clear indication that the limits of this method are reached. However, as apparent from the standard deviation of Table  \ref{tab:qbqtHR}  the hit-and-run algorithm produces  a reliable estimate for $R$ also in this case.

\subsection{General qubit-qutrit states}
\label{sec:qbqtgen}

In the most general case of qubit-qutrit states, which lie within a linear subspace of dimension $d=n^2- 1 = 35$ ($n=2\times 3=6$), see \eqref{qubitqutrit},  the ratio  according to our numerical results 
based on the hit-and-run algorithm is 
\begin{equation}
  \bar{R}^{HR}=0.026969, \quad \sigma^{HR}=0.000042,
\end{equation}
with $3.25 \times 10^8$ states. For the qubit-qutrit case a conjecture of $R=32/1199\approx 0.026688$ was made by Slater in \cite{Slaterqutrit}, which is close to the estimate obtained by us. 
Furthermore, our result is also consistent with the numerical result of  Milz et al. \cite{Milz}, i.e. $R=0.02700 \pm 0.00016$.

\subsection{Qubit--four-level qudit states}
\label{sec:qbqdgen}

For these quantum states a negative partial transpose is only a sufficient but not a necessary condition for entanglement.
All quantum states can be represented in the form

\begin{eqnarray}
\rho &=& \frac{I_8}{8} + \frac{1}{2\sqrt{2}}\sum_{i=x,y,z} \tau^{(A)}_i \,\sigma_i^{(A)} \otimes I^{(B)}_4 + \frac{1}{2\sqrt{2}}\sum^{15}_{j=1} \tau^{(B)}_j \,I^{(A)}_2 \otimes M^{(B)}_j \nonumber\\
 &+&
 \frac{1}{2\sqrt{2}}\sum_{i=x,y,z} \sum_{j=1}^{15} \nu_{i,j} \,\sigma_i^{(A)} \otimes {M}^{(B)}_j
\label{qubitqutrit}
\end{eqnarray}
with $\tau^{(A)}_i,\tau^{(B)}_j,\nu_{i,j}\in \mathbb{R}$. The basis elements ${M}^{(B)}_j$ of the four-level qudit system are considered to be identical with the basis elements of the two-qubit system, 
because $\mathbb{C}^2\otimes\mathbb{C}^2 \cong \mathbb{C}^4$.
The qubit and four-level qudit subsystems are denoted by $A$ and $B$, respectively. This means that the quantum states form a $d = 63$ dimensional convex set embedded in the $63$ dimensional linear subspace of
self-adjoint matrices with unit trace. For the ratio $R$ between PPT quantum states and all quantum state we have obtained the following numerical result
\begin{equation}
 \bar{R}^{HR}=0.001294, \quad \sigma^{HR}=0.000004
\end{equation}
with the help of the hit-and-run algorithm  with $1.2 \times 10^9$ quantum states. This numerical result is again in good agreement with the numerical result of Milz et al. \cite{Milz}.

\subsection{General qutrit-qutrit states}
\label{sec:qtqtgen}

Also for these quantum states a negative partial transpose is only a sufficient but not a necessary condition for entanglement. In this case quantum states form a $d = 80$ dimensional convex set and can 
be represented in the form

\begin{eqnarray}
\rho &=& \frac{I_9}{9} + \frac{1}{\sqrt{6}}\sum^8_{i=1} \tau^{(A)}_i \,\gamma_i^{(A)} \otimes I^{(B)}_3+ \frac{1}{\sqrt{6}}\sum_{i=1}^8 \tau^{(B)}_i \,I^{(A)}_3 \otimes \gamma^{(B)}_i  \nonumber\\
 &+&
 \frac{1}{2}\sum_{i,j=1, \dots, 8} \nu_{i,j} \,\gamma_i^{(A)} \otimes \gamma^{(B)}_j
\label{qubitqutrit}
\end{eqnarray}
with $\tau^{(A)}_i,\tau^{(B)}_i,\nu_{i,j}\in \mathbb{R}$ and the Gell-Mann matrices $\gamma_i$. For this case the hit-and-run algorithm yields the following numerical results for the ratio $R$ between PPT quantum states and all quantum states
\begin{equation}
  \bar{R}^{HR}=0.0001025, \quad \sigma^{HR}=0.0000012
\end{equation}
with a sample size containing $9 \times 10^8$ quantum states.

\section{Bell inequalities and detectable entanglement}
\label{IV}

In his seminal paper \cite{Bell} John Bell presented an inequality capturing the essence of local realistic correlations which can be violated by correlations originating from some particular  quantum states. This discovery stimulated intense research activities on Bell inequalities for different types of correlation experiments \cite{Pitowsky}. Although there are entangled quantum states, which do not violate Bell inequalities, testing for violations of these inequalities is still a
convenient tool for assessing entanglement experimentally. 
In this section we investigate 
the typicality of bipartite two-qubit entanglement which can be detected by violations of Bell inequalities.
For this purpose we focus on two types of Bell inequalities, namely
the CHSH inequality \cite{CHSH} and the inequality proposed by Collins and Gisin \cite{CG}.

\subsection{The CHSH inequality}
\label{sec:CHSH}

\begin{figure}[t!]
   \centering
  \includegraphics[width=0.4\textwidth]{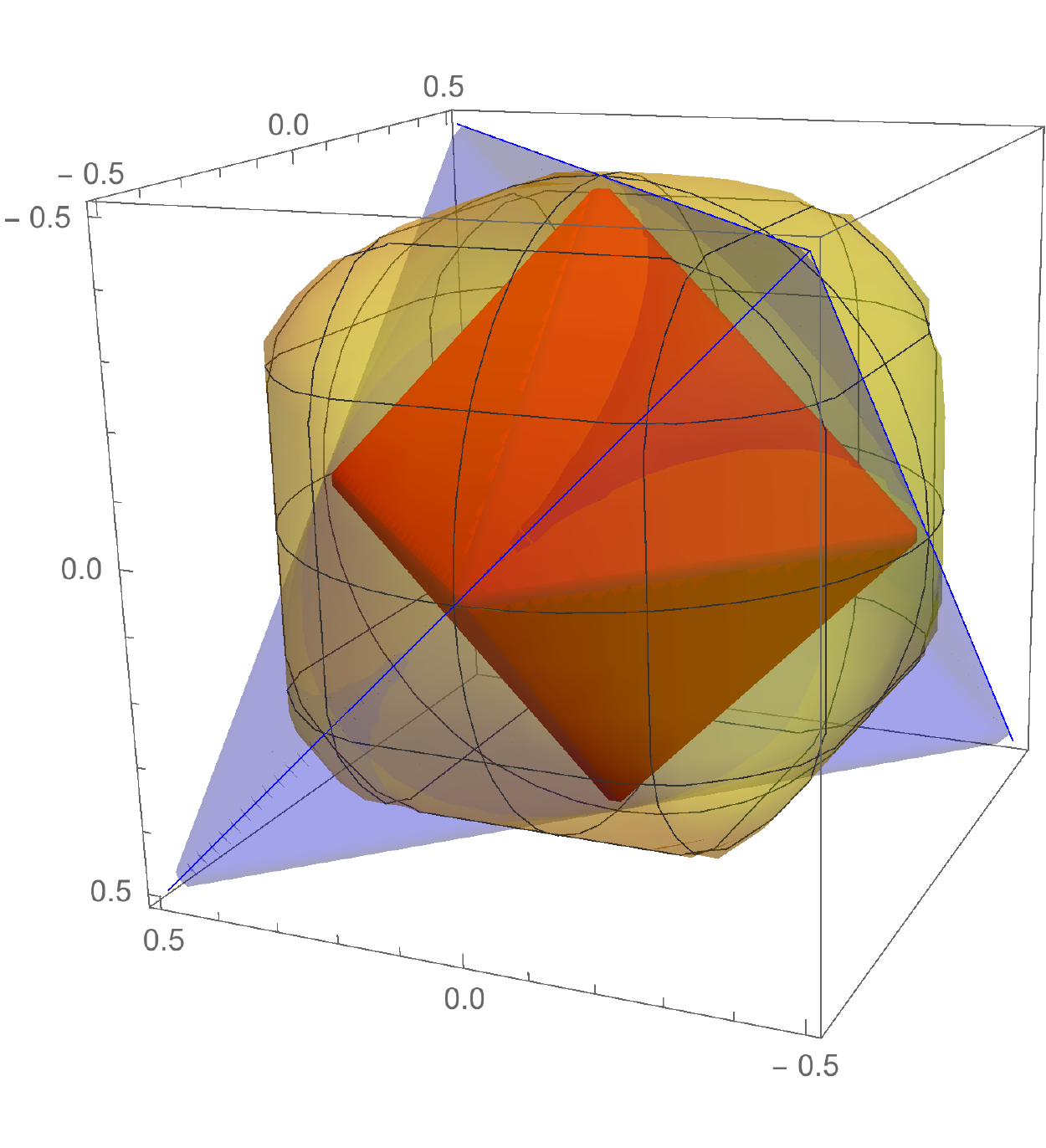}
     \caption{Schematic representation of the Steinmetz solid in \eqref{eq:Steinmetz} together with the convex set of separable Bell diagonal two-qubit states (octahedron) inside  
     the convex set of all Bell diagonal two-qubit states (tetrahedron), see also Fig. \ref{Belldiagonal}: The Steinmetz solid contains the whole octahedron and has also parts lying outside of the tetrahedron which do not represent
quantum states.}
   \label{Steinmetz}
 \end{figure}

The CHSH inequality refers to bipartite correlation experiments on sites $A$ and $B$ with an observation of two measurements on each site with two possible 
outcomes, say $\pm 1$. Therefore, four possible observables are involved, namely $A_1=\bm{v}_1 \cdot \bm{\sigma}$, $A_2=\bm{v}_2 \cdot \bm{\sigma}$, $B_1=\bm{w}_1 \cdot \bm{\sigma}$ and 
$B_2=\bm{w}_2 \cdot \bm{\sigma}$ with unit vectors $\bm{v}_1,\bm{v}_2,\bm{w}_1,\bm{w}_2$ and with the the Pauli vector $\bm{\sigma}$. An experimental setting for a particular Bell experiment is characterized by a particular quadrupel of unit vectors $(\bm{v}_1,\bm{v}_2,\bm{w}_1,\bm{w}_2)$. In terms of these observables the CHSH inequality is given by
\begin{equation}\label{CHSH}
 -2 \leqslant E(A_1B_1)+E(A_1B_2)+E(A_2B_1)-E(A_2B_2)\leqslant 2
\end{equation}
with $E(.)$ denoting the expectation value. 
Representing  a general two-qubit density matrix in the form of  \eqref{full} with the coefficients $\nu_{i,j}$ ($i,j \in \{x,y,z\}$) forming the real-valued $3 \times 3$ 
matrix $C_{\rho}$, this CHSH inequality can be written also in the equivalent form 
\begin{equation}
  -2 \leqslant 2 \left\langle \bm{v}_1, C_{\rho} \left(\bm{w}_1 +\bm{w}_2 \right) \right\rangle +  2 \left\langle \bm{v}_2, C_{\rho} \left(\bm{w}_1 -\bm{w}_2 \right) \right\rangle \leqslant 2.
\end{equation}
Following Ref. \cite{Horodecki95} and using the orthogonality of the vectors $\bm{w}_1 +\bm{w}_2$ and $\bm{w}_1 -\bm{w}_2$
 it is found that this CHSH inequality is fulfilled if and only if the  quantum state $\rho$ of \eqref{full}  fulfills the condition
\begin{equation}\label{CHSHcond}
 \lambda_1 + \lambda_2 \leqslant \frac{1}{4}
\end{equation}
with $\lambda_1$ and $\lambda_2$ denoting the two largest eigenvalues of the matrix $C^\dagger_{\rho} C_{\rho}$. Thus, for these quantum states $\rho$ no possible experimental setting of the four possible observables can cause a violation of the CHSH inequality.

In order to gain some additional insight let us consider Bell diagonal states (cf.  \eqref{eq:belldiag}). In this case Eq. \eqref{CHSHcond} yields
\begin{equation}\label{eq:Steinmetz}
 a^2_x+a^2_y \leqslant \frac{1}{4}, \quad  a^2_x+a^2_z \leqslant \frac{1}{4}, \quad  a^2_y+a^2_z \leqslant \frac{1}{4}.
\end{equation}
These three inequalities define the so-called Steinmetz solid or tricylinder which is the intersection of three cylinders of equal radii intersecting at right angles. 
As shown in Fig. \ref{Steinmetz} the Steinmetz solid 
contains not only all separable states but also some entangled states, as pointed out also in Ref. \cite{Alves} in a study of entropic inequalities.

With the help of the hit-and-run algorithm we have estimated the Euclidean volume ratio $R_{\text{CHSH}}$ of the quantum states which violate the CHSH inequality at least for one possible setting of the four 
possible observables. 
As for large samples the randomly selected points in the relevant Euclidean space become uniformly distributed this volume ratio can be estimated by the ratio of the number of points $N_{\text{CHSH}}$ in the 
Euclidean space  violating \eqref{CHSHcond} and the
total number of quantum states $N$.
In Table \ref{tab:qbqbCHSH} these
numerically determined ratios $R_{\text{CHSH}}=N_{\text{CHSH}}/N$ are shown for the different classes of two-qubit states investigated in the previous section.

\begin{table}[H]
  \centering
\begin{tabular}{|c|c|c|c|c|c|}
\hline
 & $N$  & $R_{\text{CHSH}}$ & $\sigma_{\text{CHSH}}$ \\
\hline
Bell diagonal states & $5.4 \cdot 10^9$ &  0.087021 &  0.000010\\
\hline
X-states & $4\cdot 10^9$ &  0.057276 & 0.000015 \\
\hline
Rebit-rebit states & $4\cdot 10^9$ & 0.011082 & 0.000008   \\
\hline
General two-qubit states& $4\cdot 10^9$ & 0.008221 & 0.000008 \\
\hline
\end{tabular}
\caption{
Estimates of the Euclidean volume ratios $R_{\text{CHSH}}=N_{\text{CHSH}}/N$ for all families of two-qubit quantum states discussed in Sec. \ref{III}: $N$ denotes the number of quantum states within the randomly 
selected sample and $N_{\text{CHSH}}$ is
the number of quantum states violating \eqref{CHSHcond}.}
\label{tab:qbqbCHSH}
\end{table}

Comparing the results of Table \ref{tab:qbqbCHSH} with the ratios obtained in Sec.\ref{III} it is apparent that only a small fraction of entangled states is detectable by all possible Bell experiments testing 
for a violation of the CHSH inequality. For Bell-diagonal states, for example, the Euclidean volume ratio between separable states and all quantum states is $0.5$ so that the Euclidean volume ratio between 
CHSH-detectable entangled states and all entangled states is estimated as $0.087/(1-0.5)\approx 0.174$.
The corresponding estimated ratios for
the remaining three cases of 
Table \ref{tab:qbqbCHSH} 
are given by $0.0955$ for X-states, $0.0203$ for rebit-rebit states, and $0.0108$ for general two-qubit states. These results demonstrate that most of the entangled two-qubit states are not detectable 
by CHSH-type Bell tests even if ideal measurement arrangements can be realized for all the infintely many possible measurement setups.

In practice it is impossible to perform CHSH-type Bell tests for all possible measurement arrangements. If only a finite number of Bell tests are performed the number of detectable entangled states is reduced even 
further. Thus, the natural question arises whether there is a finite list of special measurements which guarantees the detection of a large fraction of all CHSH-detectable entangled two-qubit states 
even under the assumption of ideal apparatuses. In the following we provide an answer to this question.  
In order to address this question
let us consider the simple and geometrically lucid case of Bell-diagonal quantum states. In this case the half space defined by each tangent plane of the Steinmetz solid represents a CHSH inequality for a 
particular measurement setup. 
Furthermore, the four corners of the tetrahedron characterizing all Bell-diagonal quantum states (cf.  Fig. \ref{Steinmetz}) are not inside the Steinmetz solid.
According to Table \ref{tab:qbqbCHSH}
the volume of these parts has been estimated as $8.7 \%$ of the volume of the whole tetrahedron. 
As the Steinmetz solid is formed by the intersection of three cylinders it has extreme points at which its tangent planes are not definied uniquely. In particular, it has $8$ extreme
points each of which defines three different tangent planes, one for each of the three cylinders whose intersection defines the surface of the Steinmetz solid. 
Four of these extreme points are located inside of the tetrahedron characterizing the possible quantum states (cf.  Fig. \ref{Steinmetz}). 
Therefore, it appears plausible that the three tangent planes associated with each of these four extreme points 
are capable of detecting a large fraction of all CHSH-detectable entangled states.

The corresponding $4\times 3 = 12$ inequalities defining these half spaces can
be written in the following concise way
\begin{eqnarray}
 |a_i| + |a_j| \leqslant \frac{1}{\sqrt{2}}, \quad \text{ for } i \neq j \quad \text{and}\quad  i,j \in \{x,y,z\} ,~ i \neq j. \label{Bellstra} 
\end{eqnarray}
With the hit-and-run algorithm we have estimated the Euclidean volume ratio $R_{\text{12m}}$ between Bell-diagonal states violating at least one of the $12$ inequalities \eqref{Bellstra} and all Bell-diagonal quantum states. Comparing it with the corresponding Euclidean volume ratio $R_{\text{CHSH}}$ for Bell-diagonal states we find that $86,63\%$ of all CHSH-detectable Bell-diagonal states are detected by the
$12$ inequalities \eqref{Bellstra}. Thus, these latter inequalities capture a large fraction of all CHSH-detectable entangled Bell-diagonal states.

\begin{table}[H]
\centering
\begin{tabular}{|c|c|c|c|c|c|}
\hline
 & $N$  & $R_{\text{12m}}$&$\sigma_{\text{12m}}$ & $R_{\text{12m/CHSH}}$ \\
\hline
Bell diagonal states & $5.4 \cdot 10^9$  & 0.075387 & 0.000013 & 0.8663 \\
\hline
X-states & $4\cdot 10^9$  & 0.006104 & 0.000006 & 0.1066\\
\hline
Rebit-rebit states & $4\cdot 10^9$ & 0.001766 & 0.000004 & 0.1594 \\
\hline
General two-qubit states& $4\cdot 10^9$ & 0.000044 & 0.000001 & 0.0054\\
\hline
\end{tabular}
\caption{
Estimates of the Euclidean volume ratios $R_{\text{12m}}$ 
violating at least one of the $12$ inequalities \eqref{Bellstra}
for all families of two-qubit quantum states discussed in Sec. \ref{III}: $N$ denotes the number of quantum states, 
$R_{\text{12m/CHSH}} = R_{\text{12m}}/R_{\text{CHSH}}$ is the 
fraction of all CHSH-entangled states that can be detected using one of the $12$ Bell measurements. }
\label{tab:qbqb12m}
\end{table}

Table \ref{tab:qbqb12m} summarizes numerical results for
Euclidean volume ratios $R_{\text{12m}}$ for all families of two-qubit quantum states discussed in Sec. \ref{III} which violate at least one of the $12$ inequalities \eqref{Bellstra}. From these results it is apparent that
the usefulness of these $12$ measurements for detecting entanglement quickly diminishes for non Bell-diagonal quantum states. 
In the general case, for example, only $0.54\%$ of the states which violate a CHSH inequality for some measurement setting can be detected by one of these 12 measurements.

These results clearly demonstrate that even under the assumption of ideal measurement setups only a small fraction 
$R_{\text{CHSH}}$ of the entangled states can be detected by CHSH inequalities. This fraction is reduced even further if only a finite number of measurements is taken into account in the CHSH Bell tests.
Thus, finding an inequality which is more efficient than the CHSH inequality is an important task. 
In our subsequent subsection we consider a possible candidate, the Collins-Gisin inequality \cite{CG}. It has an interesting relation to the 
CHSH inequality because there are quantum states which violate the Collins-Gisin but not the CHSH inequality and vice versa. 
Because our approach allows to quantify the efficiency of Bell inequalities we are able to compare quantitatively these two families of inequalities.

\subsection{Collins-Gisin inequality}
\label{sec:CG}

For the case of three possible measurements on both sites $A$ and $B$, each of which has two possible oucomes, Collins and Gisin   \cite{CG} proposed new inequalities based on results of Pitowsky and Svozil \cite{PSvozil}. 
Apart from 
variations of the CHSH inequality these inequalities also involve a new one, namely
\begin{eqnarray} \label{CG}
 0\leqslant 4 +  E(A_1) + E(A_2) + E(B_1)+ E(B_2)+  E(A_1B_1)+ E(A_1B_2)  \\
+E(A_2B_1)+ E(A_2B_2)+ E(A_1B_3)+ E(A_3B_1)-E(A_2B_3)-E(A_3B_2). \nonumber
\end{eqnarray}
Using the general representation \eqref{full} of two-qubit quantum states this inequality can be written in the equivalent form

\begin{eqnarray}\label{CGrew}
 0\leqslant 2+  \langle \bm{v}_1 +\bm{v}_2, \bm{\tau}^{(A)}_{\rho} \rangle +  \langle \bm{w}_1 +\bm{w}_2, \bm{\tau}^{(B)}_{\rho} \rangle + \left\langle \bm{v}_1, C_{\rho} \left(\bm{w}_1 +\bm{w}_2 +\bm{w}_3 \right) \right\rangle 
 \nonumber \\
 + \left\langle \bm{v}_2, C_{\rho} \left(\bm{w}_1 +\bm{w}_2 -\bm{w}_3 \right) \right\rangle + \left\langle \bm{v}_3, C_{\rho} \left(\bm{w}_1 -\bm{w}_2 \right) \right\rangle.
\end{eqnarray}
In addition to the matrix
$C_{\rho}$, which appears also in the CHSH inequality,  this inequality contains  the $6$ parameters  $\bm{\tau}^{(A)}_{\rho}=(\tau^{(A)}_x,\tau^{(A)}_y,\tau^{(A)}_z)^T$ and 
$\bm{\tau}^{(B)}_{\rho}=(\tau^{(B)}_x,\tau^{(B)}_y,\tau^{(B)}_z)^T$ ($T$ denotes the transposition) characterizing the quantum state $\rho$. 

Analogous to our previous discussion of the CHSH inequality, which led to condition \eqref{CHSHcond}, we are interested in determining the minimum of the right hand side of this inequality with respect to  all possible measurement settings. This minimum defines the surface 
of a convex body. Quantum states lying inside this body are not able to violate any kind of Collins-Gisin type inequality. By applying the Cauchy-Bunyakovsky-Schwarz inequality
\begin{equation}
 |\langle x,y \rangle| \leqslant \|x\| \|y\|, \quad x,y \in \mathbb{C}^n \nonumber
\end{equation}
we can minimize over all unit vectors $\bm{v}_1$, $\bm{v}_2$ and $\bm{v}_3$. This minimization yields the inequality

\begin{eqnarray}
 &&2 +  \langle \bm{w}_1 +\bm{w}_2, \bm{\tau}^{(B)}_{\rho} \rangle + \left\langle \bm{v}_1, C_{\rho} \left(\bm{w}_1 +\bm{w}_2 +\bm{w}_3 \right) + \bm{\tau}^{(A)}_{\rho} \right\rangle  \nonumber \\ 
 &&+\left\langle \bm{v}_2, C_{\rho} \left(\bm{w}_1 +\bm{w}_2 -\bm{w}_3 \right) + \bm{\tau}^{(A)}_{\rho} \right\rangle 
 + \left\langle \bm{v}_3, C_{\rho} \left(\bm{w}_1 -\bm{w}_2 \right) \right\rangle \geqslant \nonumber \\ 
  &&\geqslant 2 +  \langle \bm{w}_1 +\bm{w}_2, \bm{\tau}^{(B)}_{\rho} \rangle - \|C_{\rho} \left(\bm{w}_1 +\bm{w}_2 +\bm{w}_3 \right) + \bm{\tau}^{(A)}_{\rho}\| \nonumber \\
  &&-\|C_{\rho} \left(\bm{w}_1 +\bm{w}_2 -\bm{w}_3 \right) + \bm{\tau}^{(A)}_{\rho}\|
 -\|C_{\rho} \left(\bm{w}_1 -\bm{w}_2\right)\| \label{CGnum}
\end{eqnarray}
with equality holding if and only if the scalar products on the left hand side are as small as possible.

The inequality \eqref{CGnum} can be further minimized in the case of Bell diagonal states for which $\bm{\tau}^{(A)}_{\rho}=\bm{0}$ and $\bm{\tau}^{(B)}_{\rho}=\bm{0}$ by maximizing the quantity
\begin{equation}
 \|C_{\rho} \left(\bm{w}_1 +\bm{w}_2 +\bm{w}_3 \right)\|+\|C_{\rho} \left(\bm{w}_1 +\bm{w}_2 -\bm{w}_3 \right)\|+\|C_{\rho} \left(\bm{w}_1 -\bm{w}_2\right)\| \label{eq:tomax}
\end{equation}
for all unit vectors $\bm{w}_1$, $\bm{w}_2$ and $\bm{w}_3$. 
Applying the polarization identity we obtain the relations
\begin{eqnarray}
 \|C_{\rho} \left(\bm{w}_1 +\bm{w}_2 +\bm{w}_3 \right)\|&=& \sqrt{\|C_{\rho} \left(\bm{w}_1 + \bm{w}_2 \right)\|^2+\|C_{\rho} \bm{w}_3\|^2+2 \langle C_{\rho} \left(\bm{w}_1 + \bm{w}_2 \right), C_{\rho} \bm{w}_3 
 \rangle },\nonumber \\
  \|C_{\rho} \left(\bm{w}_1 +\bm{w}_2 -\bm{w}_3 \right)\|&=& \sqrt{\|C_{\rho} \left(\bm{w}_1 + \bm{w}_2 \right)\|^2+\|C_{\rho} \bm{w}_3\|^2-2 \langle C_{\rho} \left(\bm{w}_1 + \bm{w}_2 \right), C_{\rho} \bm{w}_3 
 \rangle}. \nonumber 
\end{eqnarray}
Therefore, the maximum of 
\begin{equation}
 \|C_{\rho} \left(\bm{w}_1 +\bm{w}_2 +\bm{w}_3 \right)\|+ \|C_{\rho} \left(\bm{w}_1 +\bm{w}_2 -\bm{w}_3 \right)\| \nonumber
\end{equation}
is obtained if and only if $C_{\rho} \left(\bm{w}_1+\bm{w}_2 \right) \perp C_{\rho} \bm{w}_3$. 
As $\bm{w}_1 +\bm{w}_2 \perp \bm{w}_1 -\bm{w}_2$ in this case we can parametrize the unit vectors in terms of an angle $\alpha \in [0,\pi/2]$ and in terms of two mutually orthogonal unit vectors $\bm{c}$ and $\bm{c}'$ in the form
\begin{eqnarray}
 \bm{w}_1 +\bm{w}_2&=& 2 \bm{c} \cos \alpha, \nonumber \\
 \bm{w}_1 -\bm{w}_2&=&2 \bm{c}'\sin \alpha, \nonumber \\
 \bm{w}_1 +\bm{w}_2 &\perp& \bm{w}_3. \nonumber 
\end{eqnarray}
In terms of this parametrization relation \eqref{eq:tomax} reduces to the form
\begin{equation}
 2 \sqrt{4 \|C_{\rho} \bm{c}\|^2 \cos^2 \alpha + \|C_{\rho} \bm{w}_3\|^2}+2 \sin \alpha \|C_{\rho} \bm{c}'\|, \nonumber
\end{equation}
and its maximum with respect to the angle $\alpha$ is given by
\begin{equation}
 \frac{\sqrt{\left(4\|C_{\rho} \bm{c}\|^2+\|C_{\rho} \bm{c}'\|^2\right)\left(4\|C_{\rho} \bm{c}\|^2+\|C_{\rho} \bm{w}_3\|^2\right)}}{\|C_{\rho} \bm{c}\|}. \nonumber
\end{equation}

\begin{figure}[t!]
   \centering
     \includegraphics[width=0.4\textwidth]{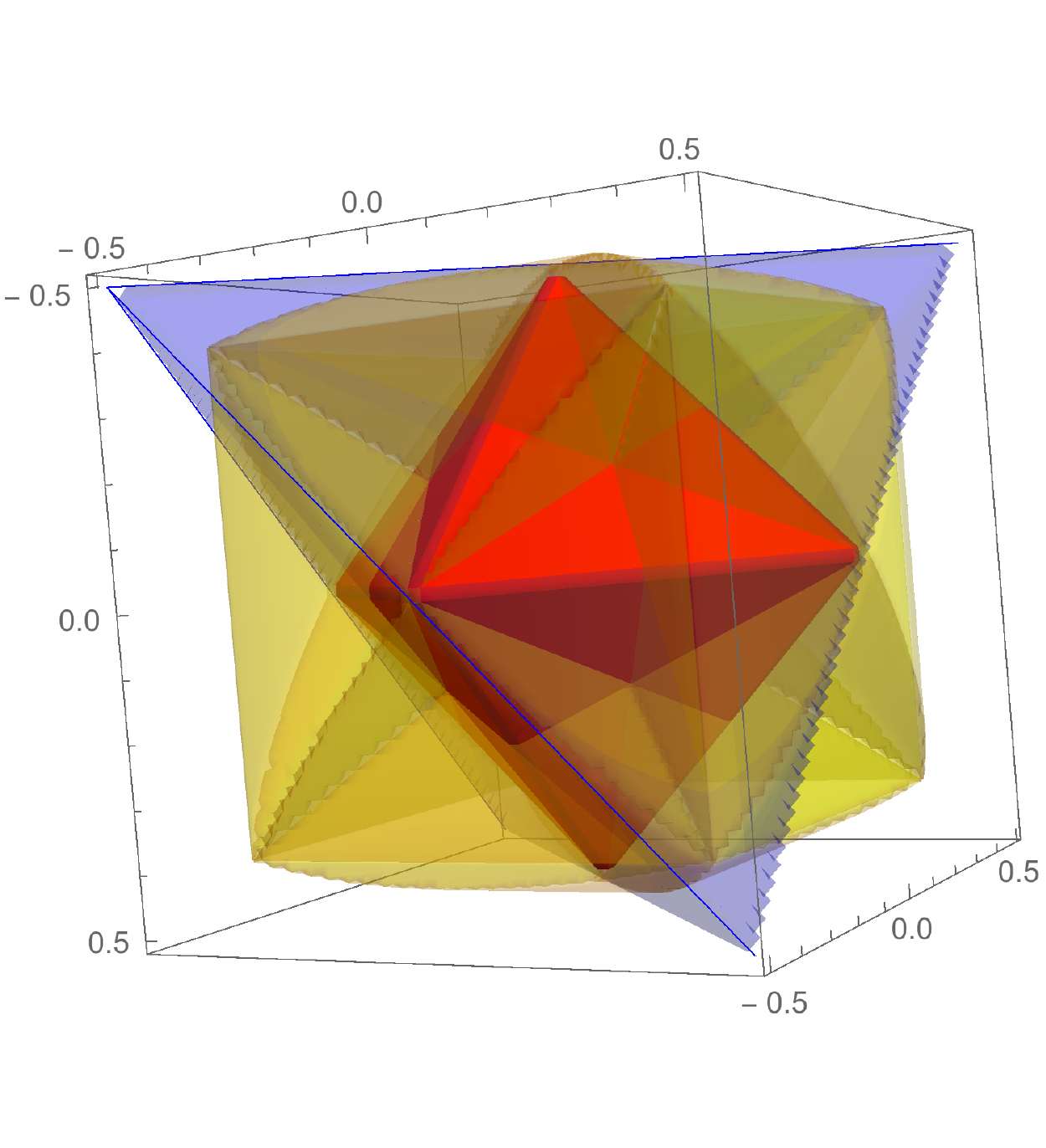}
     \caption{Schematic representation of the convex set defined by the inequalities in \eqref{eq:CGbody} together with the convex set of separable Bell diagonal two-qubit states (octahedron) inside  
     the convex set of all Bell diagonal two-qubit states (tetrahedron), ( cf. Figs. \ref{Belldiagonal} and \ref{Steinmetz}): Points lying outside of the tetrahedron do not represent quantum states.}
   \label{CGbody}
 \end{figure}

For Bell diagonal states $C^T_{\rho}C_{\rho}=\operatorname{diag}(a^2_x, a^2_y, a^2_z)$ and the maximum of this expression is achieved if and only if $\bm{w}_3 \parallel \bm{c}'$ and if both vectors are the eigenvectors of the second 
largest eigenvalue of $C^T_{\rho}C_{\rho}$.
So  \eqref{eq:tomax} simplifies to
\begin{equation}
 \frac{4\|C_{\rho} \bm{c}\|^2+\|C_{\rho} \bm{c}'\|^2}{\|C_{\rho} \bm{c}\|}, \nonumber
\end{equation}
which results in the $6$ inequalities
\begin{equation}\label{eq:CGbody}
 0\leqslant 2 - \frac{4 a^2_i + a^2_j}{|a_i|}, \quad a^2_i \geqslant a^2_j \geqslant a^2_k
\end{equation}
with $i,j,k \in \{x,y,z\}$. These inequalities define a convex body, which is larger than the Steinmetz solid obtained for the CHSH type inequalities. As shown
in Fig. \ref{CGbody} this convex body contains some entangled and all separable two-qubit quantum states.

With the help of the hit-and-run algorithm the volume ratio $R_{\text{CG}}$ = $N_{\text{CG}}/N =  0.03677 \pm 0.00001$ has been estimated with $N$ denoting all Bell-diagonal quantum states and $N_{CG}$ 
denoting the number of all detectable entangled Bell-diagonal
quantum states which do not fulfill \eqref{eq:CGbody}. Comparing this result with the corresponding ratio of the CHSH inequality, i.e.
$R_{\text{CHSH}}=0.08702 \pm 0.00001$, it is apparent that for Bell-diagonal two-qubit quantum states
the CHSH inequality can detect entanglement more efficiently than the Collins-Gisin inequality. However, it turns out this property is not valid for arbitrary two-qubit states of the $15$ dimensional 
Euclidean vector space. This may be traced back to the fact that the information on the density matrix $\rho$ stored in the vectors $\bm{\tau}^{(A)}_{\rho}$ and $\bm{\tau}^{(B)}_{\rho}$ is exploited by the Collins-Gisin inequality efficiently,
while the CHSH inequality does not take this information into account at all. 
 
\begin{figure}[t!]
   \centering
   \includegraphics[width=0.9\textwidth]{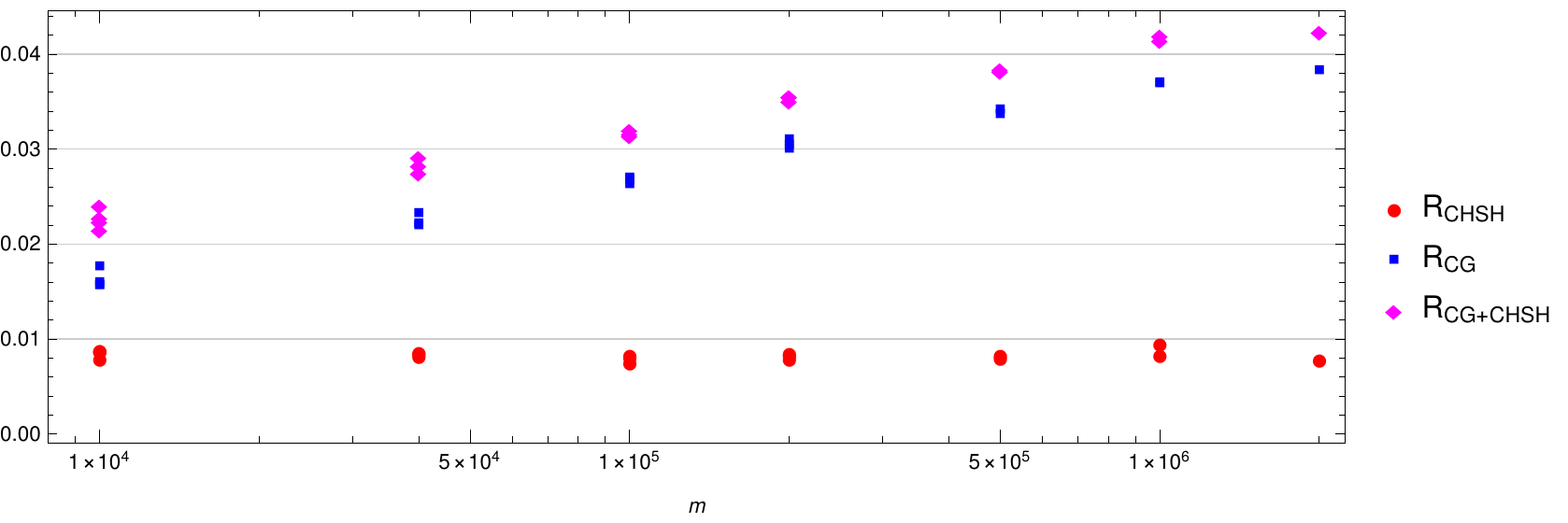}
     \caption{Fraction of two-qubit quantum states violating the Collins-Gisin inequality $R_{\text{CG}}$, the CHSH inequality $R_{\text{CHSH}}$ and either one or the other $R_{\text{CG}+\text{CHSH}}$: For each point $m$ random measurement settings are generated and $10^6$ states are tested for possible violations.}
   \label{fig:CGrandom}
 \end{figure}

With the help of the hit-and-run algorithm violations of the Collins-Gisin inequality have been investigated
for general two-qubit quantum states using sets of random measurements.
As this procedure is very time consuming, only $10^6$ points have been generated per run. 
In Fig. \ref{fig:CGrandom} the fraction of two-qubit quantum states violating the Collins-Gisin inequality $R_{\text{CG}}$, violating the CHSH inequality $R_{\text{CHSH}}$ and violating either the one or the other $R_{\text{CG}+\text{CHSH}}$ are shown for different numbers of randomly selected measurements. Apparently the combination of both inequalities results in the highest ratios, which is consistent with former results \cite{CG}.
This is due to the fact that there are states which violate one of these inequalities but not the other one. These numerical results demonstrate convincingly that as far as arbitrary two-qubit quantum states are concerned the Collins-Gisin inequality is capable of detecting entanglement more efficiently than the CHSH inequality. According to Fig.  \ref{fig:CGrandom} there is no convincing convergence of our numerical results with increasing numbers of measurements even at a level of $2\cdot 10^6$ random measurement settings. Therefore,
we have investigated for given quantum states the right hand side of inequality \eqref{CGnum}, which is already optimized for $3$ vectors.
This way we have obtained the following estimate
\begin{equation} \label{CGest}
 R_{\text{CG}}=0.07128 \pm 0.00002.
\end{equation}
Comparison of this result with the corresponding results of the CHSH inequality, i.e. 
 $R_{\text{CHSH}}=0.008221\pm 0.000008$ (cf. Table \ref{tab:qbqbCHSH}), also hints at the better performance of the Collins-Gisin inequality as far as detectable entanglement of two-qubit quantum states is concerned. 
Imposing the condition of either violating
\eqref{CHSHcond} or \eqref{CGrew} we obtain the estimate
\begin{equation} \label{CG+CHSHest}
 R_{\text{CG}+\text{CHSH}}= 0.073364 \pm 0.000021.
\end{equation}
Although this demonstrates an improvement in the ratio of detectable entanglement it should be kept in mind that still $90.3 \%$ of all entangled two-qubit quantum states remain undetected by these Bell inequalities.

\section{Summary and conclusions}
\label{V}

We have investigated the Euclidean volume ratios $R$ between PPT and all quantum states in several bipartite quantum systems. For this purpose a new approach has been developed. On the analytical side it is based 
on the 
Peres-Horodecki criterion and tools involving Newton identities and Descartes' rule of signs and on the numerical side it involves two numerical 
methods based on the multiphase Monte Carlo method combined with the Muller method and on the hit-and-run algorithm. 
 
For two-qubit states we have been able to estimate this Euclidean volume ratios $R$ with high accuracy in several interesting cases. 
Thereby, the analytically obtainable volume ratio of two-qubit Bell diagonal states, i.e., $R=0.5$, of X-states, i.e., $R=0.4$ \cite{Milz}, and of rebit-rebit states, i.e., $R=\frac{29}{64}$ \cite{Lovas},
have been used as a benchmark to test the numerical accuracy and characteristic properties of the Monte Carlo methods used in our numerical approach. For general two-qubit states our results of the 
multiphase Monte Carlo method, i.e. $R=0.243 \pm 0.007$, and the hit-and-run algorithm, i.e. $R=0.242444 \pm 0.000027$
are close to the recent analytical and numerical results of Slater and Dunkl \cite{Slater3} supporting the conjecture that $R=8/33\approx 0,24242$, and are consistent with the numerical result of 
Shang et al. \cite{Shang}, i.e. $R=0.242 \pm 0.002$, Milz et al. \cite{Milz}, i.e., $R = 0.24262 \pm 0.0134$, and Fei et al. \cite{Fei}, i.e., $R=0.24243 \pm 0.00001$. Compared to other numerical approaches 
these accuracies can already be achieved with  significantly lower sample sizes.

We have demonstrated that already in qubit-qutrit systems  the advantage of the Muller and Mutliphase Monte Carlo method, namely generating quantum states uniformly, is compromised 
by increasing the dimension of the Euclidean space from $d=3$ (for Bell-diagonal qubit-qubit systems) to $d=35$ (for general qubit-qutrit systems).

Our numerical investigations demonstrate that already for $d=24$ the multiphase Monte Carlo approach requires
large numbers of points in order to find at least some quantum states. On the 
other hand in the hit-and-run algorithm quantum states are not generated uniformly and uniform distributions are obtained only in the limit of large sample sizes.
For general qubit-qutrit states, where all PPT quantum states are separable, our result of the hit-and-run algorithm,
i.e., $R=0.026969 \pm 0.000042$, is again consistent with the conjecture of Slater, i.e., $R=32/1199\approx 0.026688$ \cite{Slaterqutrit}, and with the numerical result of  Milz et al. \cite{Milz}, 
i.e., $R=0.02700 \pm 0.00016$. We have also tested our approach for a qubit-qudit system with a four-level qudit ($d=63$) and for qutrit-qutrit systems ($d=80$). 
For this particular qubit-qudit system we have obtained the result $R=0.001294 \pm 0.000004$ which is consistent with the result of
Milz et al. \cite{Milz}. 
As a new result of this approach we find the ratio $R=0.0001025 \pm 0.0000012$ for general qutrit-qutrit quantum states.

With the help of our numerical approach we have also investigated  the typicality of detectable bipartite entanglement in two-qubit systems which can be detected by violations of Bell inequalities.
Our results demonstrate that for general two-qubit quantum states the Collins-Gisin type Bell inequality is capable of detecting more entangled states than the CHSH inequality.  
Whereas the CHSH type Bell inequality can detect only $1\%$ of all entangled two-qubit states, the Collin-Gisin inequality is capable of detecting 
 almost $9.4\%$ of all entangled two-qubit states.  A combined test of both inequalities is even capable of detecting $9.6\%$ of all entangled two-qubit states. 
 
 For the special case of Bell-diagonal two-qubit quantum states we have also presented an analytical criterion for violating the Collins-Gisin inequality at least for one possible measurement setup. Within this special class of quantum states this analytical result generalizes the result of Horodecki et al.
 \cite{Horodecki95} (cf. \eqref{CHSHcond}) for the CHSH inequality to the Collins-Gisin inequality. Contrary to the case of general two-qubit quantum states it turned out that for Bell-diagonal two-qubit quantum states the CHSH inequality is more efficient in detecting entanglement than the Collins-Gisin inequality. 
As these results apply to the highly idealized situation, in which Bell tests can be realized 
with all possible measurement setups, we have also addressed the question which finite number of Bell measurements is capable of detecting a large part of  entangled states. 
For the CHSH inequality we have proposed a list of $12$ special measurement setups. Despite their small number these measurement setups are capable of detecting already $86.63\%$ of all detectable Bell-diagonal 
entangled two-qubit states. 

All our numerical results support the expectation that
the Euclidean volume ratios between PPT quantum states and all quantum states in bipartite quantum systems is decreasing fast and tends to zero with increasing dimension of the quantum systems involved \cite{infinite}. 
Despite the resulting dominance of entangled bipartite quantum states   with negative partial transpose we have demonstrated quantitatively that already in two-qubit systems the detectability of entanglement 
by Bell inequalities is very limited. Therefore, this dichotomy between the abundance of entangled quantum states on the one hand and the detectability of entanglement by Bell-type inequalities on the other hand deserves further investigation.

\ack
The authors acknowledge stimulating discussions with A.R.P. Rau. This research was supported by the Deutsche 
Forschungsgemeinschaft (DFG) -- SFB 1119 -- 236615297 and by the 
DFG 
under Germany's Excellence Strategy -- Cluster of Excellence Matter and Light for Quantum Computing (ML4Q) EXC 2004/1 -- 390534769.

\end{document}